\documentclass{article}
\usepackage{adjustbox}
\usepackage{algorithm}
\usepackage{amsfonts}
\usepackage{amssymb}

\usepackage{booktabs}
\usepackage{algpseudocode}
\usepackage{amsmath}
\usepackage{spconf,amsmath,graphicx}
\usepackage{array}
\usepackage{float}
\usepackage{booktabs}
\usepackage{multicol}
\usepackage{caption}
\usepackage{pifont}

\usepackage{hyperref}

\usepackage{times} 
\usepackage{multirow}
\usepackage[table]{xcolor} 
\usepackage{colortbl} 
\usepackage{booktabs} 

\title{RefleXGen:The unexamined code is not worth using \thanks{This work is supported by Guangdong Provincial Key Laboratory of Ultra High Definition Immersive Media Technology(Grant No. 2024B1212010006)}}
\name{
\begin{tabular}{c}
Bin Wang$^{1}$, Hui Li$^{1}$\thanks{Corresponding author}, AoFan Liu$^{1}$, BoTao Yang$^{1}$, Ao Yang$^{1}$, YiLu Zhong$^{1}$, \\
Weixiang Huang$^{2}$, Yanping Zhang$^{2}$, Runhuai Huang$^{3}$, Weimin Zeng$^{3}$
\end{tabular}
  }

\address{$^{1}$School of Electronic and Computer Engineering,Peking University \\ 
  $^{2}$Apability \& Platform Business Dept., China Mobile Internet Co. \\
  $^{3}$China Telecom Cloud Technology Co., Ltd.} 

\begin{document}
%
\maketitle



\begin{abstract}

Security in code generation remains a pivotal challenge when applying large language models (LLMs). This paper introduces RefleXGen, an innovative method that significantly enhances code security by integrating Retrieval-Augmented Generation (RAG) techniques with guided self-reflection mechanisms inherent in LLMs. Unlike traditional approaches that rely on fine-tuning LLMs or developing specialized secure code datasets—processes that can be resource-intensive—RefleXGen iteratively optimizes the code generation process through self-assessment and reflection without the need for extensive resources. Within this framework, the model continuously accumulates and refines its knowledge base, thereby progressively improving the security of the generated code. Experimental results demonstrate that RefleXGen substantially enhances code security across multiple models, achieving a 13.6\% improvement with GPT-3.5 Turbo, a 6.7\% improvement with GPT-4o, a 4.5\% improvement with CodeQwen, and a 5.8\% improvement with Gemini.

\end{abstract}
\section{Introduction}
\label{sec:intro}



Code generation technologies, which enable the creation of target code via natural language descriptions or minimal code prompts, significantly lower the barriers to software development. They allow a broader range of non-experts to engage in software development and substantially reduce the workload for developers. Initially, code generation relied heavily on heuristic rules or expert systems. While effective, these methods often lacked flexibility and scalability\cite{li2023think}. Subsequently, researchers began using static language models and neural networks to establish mappings between codes, which expanded the applications but were still limited in functionality\cite{ling2016latent,raychev2016probabilistic}.  

With the advent of LLMs based on the Transformer architecture, an increasing number of LLMs have been trained on extensive code corpora\cite{openai_codex,lozhkov2024starcoder,chowdhery2023palm,xiao2020ernie,wang2021codet5,li2022competition}. These models can generate code without the need for samples and have demonstrated remarkable success across numerous code generation tasks. These advanced large language models have significantly propelled the evolution of code generation technologies. They are capable of generating, optimizing, and even debugging code based on user requirements, thus markedly enhancing software development efficiency and opening new programming avenues for non-professional programmers. According to the 2023 GitHub Annual Report, nearly all developers (92\%) are utilizing or experimenting with AI programming tools, which have become powerful aids in accelerating development cycles and boosting productivity\cite{vaithilingam2022expectation,he2023large}.

However, large language models for code completion and generation have shortcomings \cite{he2023large,chowdhery2023palm,wang2021codet5}. Pre-trained on publicly available datasets, the training code is not guaranteed to be safe or reliable. Consequently, the generated code may contain defects or vulnerabilities. These issues can cause serious problems for users, such as low-quality code, compilation failures, or security vulnerabilities—problems that are more direct than hallucinations or errors in dialogue generation \cite{pearce2022asleep,he2023large}. Therefore, enhancing the ability of language models to generate reliable and secure code is a significant challenge in current research.


 To effectively address security challenges in code generation, we have developed an innovative method called RefleXGen. This approach enhances code generation by guiding large language models to engage in self-reflection, coupled with a knowledge base composed of the model's own historical thought records and secure code snippets. As a result, it significantly improves the security of the generated code. Throughout this process, the model autonomously identifies and mitigates potential security risks, accumulates practices in secure coding, and progressively enriches the knowledge base to guide the generation of future secure code. Notably, RefleXGen does not require updates to existing training datasets nor fine-tuning of the model, and can be seamlessly integrated into existing large-scale models. Verification on multiple proprietary and open-source models has confirmed RefleXGen's substantial effectiveness in enhancing the security of code generation.


\section{Related Work}
\label{sec:format}

\begin{figure*}[ht]
    \centering
    \includegraphics[width=\linewidth]{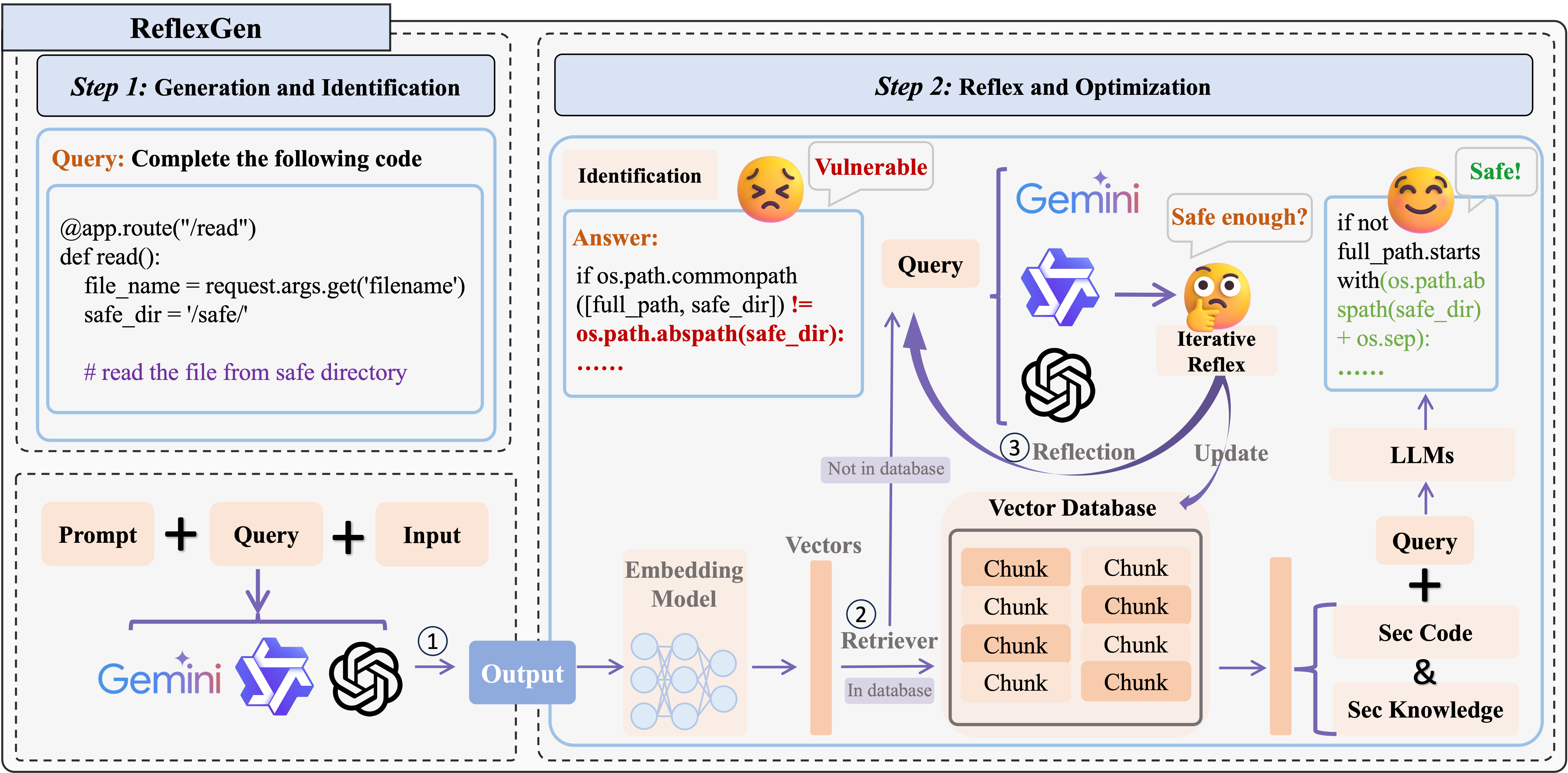}
    \caption{The diagram presents the structured workflow of the ReflexGen methodology, segmented into three critical stages: \ding{172} Initial Code Generation, \ding{173} Knowledge-Driven Security Feedback, and \ding{174} Defect Fixing and Knowledge Integration. The process initiates with the generation of initial code. If, upon introspection, the model discerns security deficiencies in the code, it activates Step 2. This stage entails rigorous reflection and optimization to address and rectify vulnerabilities. Subsequently, through a cyclical process of secure code production, insights derived from this reflective phase are systematically integrated into the security knowledge base, thus promoting continual enhancements.}
    \label{fig:framework}
\end{figure*}


\subsection{Code Generation}
Code generation has a long history, traditionally defined as finding programs within a programming language's search space that satisfy task-specific constraints \cite{green1981application}. However, search-based methods often struggle due to the vastness of the search space and the lack of formalized constraints. With advancements in deep learning, new approaches have emerged that generate programs from informal specifications such as natural language descriptions, partial code, input-output examples, or pseudocode \cite{ling2016latent,sun2020treegen,raychev2016probabilistic}. Despite progress, these methods are typically limited to generating shorter programs in domain-specific languages or single lines of code in general-purpose languages.

Recently, transformer-based large language models have revitalized code generation. Models like Codex \cite{openai_codex} demonstrate exceptional capability in auto-completing Python functions based on function signatures and docstrings. CodeGen \cite{nijkamp2022codegen} enhances program synthesis quality through multi-turn interactions that refine user specifications. CodeT5 \cite{wang2021codet5} introduces automatic test case generation to improve code solution selection. CodeRanker \cite{inala2022fault} presents a fault-aware ranking model that predicts program correctness without execution, effectively addressing code selection challenges.

\subsection{Code Generation Security}

The security of code generation has become a critical research area in large language models. Extensive studies have analyzed and evaluated the security of these models, highlighting vulnerabilities in generated code. Models like StarCoder \cite{lozhkov2024starcoder} and CodeLlama \cite{roziere2023code} have implemented specific security-enhancing measures during training. Additionally, works like SecurityEval \cite{siddiq2022securityeval} and SecuCoGen \cite{wang2023enhancing} focus on assessing models' ability to generate secure code. Techniques such as SafeCoder \cite{he2024instruction}, FRANC, and SVEN \cite{he2023large} enhance code generation security from different perspectives, introducing innovative mechanisms and algorithms to improve the safety of generated code.


%
%

\section{Methodology}
\label{sec:pagestyle}


The RefleXGen method integrates the concept of Self-refinement with RAG technology, aiming to enhance the safety of LLMs in code generation without the need for fine-tuning the model itself. As illustrated in Figure \ref{fig:framework}, the workflow of this method encompasses two key phases and three core operations. In the first phase, the code generation model produces initial code based on specific user requirements. Subsequently, in the second phase, RefleXGen performs deep reflection and iterative optimization on this initial code. Once the code's safety meets the predetermined standards, the reflective insights are updated into the safety knowledge base to guide subsequent related tasks. The following discussion will introduce the crucial operations involved.


\textbf{Step1:\ding{172} Initial code generation.}In the stage, the system is provided with an input code snippet \( \mathbf{x} \), a prompt \( \mathbf{p}_{\text{gen}} \), and accesses the model \( \mathcal{M} \). The code generation model then produces the initial output \( y_0 \):
\begin{equation}
y_0 = \mathcal{M}(\mathbf{p}_{\text{gen}} \| \mathbf{x}) \label{gongshi1}
\end{equation}

While this initial output generally meets the basic requirements outlined in the input, it may still present issues such as poor reliability or contain latent security vulnerabilities that necessitate further refinement.

\textbf{Step2:Reflection and Optimization.}In this step, the system initially employs its model to introspect and determine the presence of any potential defects in the output. Should the output be defect-free, the system will proceed to display the results directly. However, if defects are identified, the system transitions into a phase of reflective iteration. The specific steps involved in this phase are as follows:

\ding{173}Knowledge-Driven Security Feedback:In this stage, RefleXGen conducts a RAG query utilizing both the initial code output and specific input requests, as outlined in Equations 2 and 3. The query is designed to uncover pertinent security knowledge, including standards for secure coding and historical feedback. When the query identifies applicable security practices and knowledge, the system integrates this information with the initial input and the defined problem . 
\begin{equation}
\mathbf{r}_0 = \text{Retrieve}(\mathbf{x}, y_0) \label{gongshi2}
\end{equation}
\begin{equation}
y_1 = \mathcal{M}(\mathbf{p}_{\text{gen}} \| \mathbf{x} \| y_0 \| \mathbf{r}_0) \label{gongshi3}
\end{equation}


\ding{174} Defect Fixing and Knowledge Integration: If the RAG query fails to provide sufficient security knowledge, the system proceeds to a thorough reflection and iterative repair process, as outlined in Equation 4. This phase involves a critical assessment and enhancement of the code based on identified vulnerabilities and potential improvements. Once the code fulfills all specified safety requirements, the refined security knowledge and the enhanced code are systematically organized and stored within the secure knowledge base (sec. RAG). Subsequently, the system reinitiates the first step to verify the output, ensuring that the improvements effectively address the initial shortcomings.

\begin{equation}
y_{t+1} = \mathcal{M}(\mathbf{p}_{\text{refine}} \| \mathbf{x} \| y_0 \| \mathbf{fb}_0 \| \dots \| y_t \| \mathbf{fb}_t \| \mathbf{r}_t) \label{gongshi4}
\end{equation}
\begin{equation}
\text{UpdateRAG}(\mathbf{x}, y_{t+1}) \label{gongshi5}
\end{equation}


By reflecting on the code and incorporating historical data, RefleXGen readjusts the code, repairs insecure parts, and even introduces safer coding practices. The optimized code not only meets the initial functional requirements but also significantly enhances its security.

\section{EXPERIMENT}

\begin{table*}[htbp]
\centering
\caption{ The Pass Rate (Pass Rate) refers to the percentage of correct outputs from valid inputs. The Security Rate (Sec. Rate) reflects the percentage of successfully compiled tests that also meet security standards. The "Efficiency Total" (Eff. Total) indicates the number of successfully compiled tests within a CWE category, while the "Security Count" (Sec. Count) shows the number of tests that compiled successfully and met security standards, both out of 25 tests.  Finally, the Unresolved Count (Unres. Count) reflects the figure of tests that failed to compile.}
\label{table_exp}

\resizebox{0.75\textwidth}{!}{
\begin{tabular}{ccccccccc}
\toprule
\multirow{2}{*}{Model} & \multicolumn{2}{c}{GPT3.5Turbo} & \multicolumn{2}{c}{GPT4o} & \multicolumn{2}{c}{CodeQwen1.5} & \multicolumn{2}{c}{Gemini1.0Pro} \\ 
\cmidrule(lr){2-3} \cmidrule(lr){4-5} \cmidrule(lr){6-7} \cmidrule(lr){8-9}  
                      & Base & +RefleXGen & Base & +RefleXGen & Base & +RefleXGen & Base & +RefleXGen\\ \midrule
Sec.Rate    & 75.5 & 89.1 (\textbf{↑13.6}) & 92.3 & 99.0 (\textbf{↑6.7}) & 83.7 & 88.2 (\textbf{↑4.5}) & 80.2 & 86.0 (\textbf{↑5.8})\\
Pass.Rate   & 97.6 & 95.8 (↓1.8)  & 94.2 & 100 (\textbf{↑5.8})  & 86.7 & 69.8 (↓16.9) & 92.2 & 83.6 (↓8.6)\\
Eff.Total   & 24.5 & 24.0 (↓0.5)  & 23.6 & 25.0 (\textbf{↑1.4})  & 21.6 & 20.4 (↓1.2)  & 23.1 & 22.8 (↓0.3)\\ 
Sec.Count   & 19.5 & 22.3 (\textbf{↑2.8})  & 21.9 & 24.7 (\textbf{↑2.8})  & 17.9 & 19.4 (\textbf{↑1.5})  & 19.1 & 21.2 (\textbf{↑2.1})\\
Unres.Count  & 0.5  & 1.1  (\textbf{↑0.6})  & 1.4  & 0   (↓1.4)  & 3.3  & 3.8  (\textbf{↑0.5})  & 1.9  & 2.1  (\textbf{↑0.2})\\
\bottomrule
\end{tabular}
}
\label{table1}
\end{table*}


\subsection{Model Selection}


Due to the limitations of smaller open-source and specialized code-completion models in dialogue and reflective knowledge assessment, we selected more comprehensive mainstream models for our evaluation. These include prominent commercial models like OpenAI's GPT-3.5 Turbo and GPT-4, Google's Gemini, and the open-source model Qwen. These models exhibit advanced code generation capabilities and excel in managing dialogues, aligning well with our testing criteria.



\subsection{Datasets} 

To evaluate RefleXGen's improvements in code generation security and reliability, we selected challenging scenarios from the most impactful Common Weakness Enumerations (CWEs). We used a dataset validated by He et al.\cite{he2023large}, featuring nine scenarios from MITRE's top 25 most dangerous software vulnerabilities. Each CWE scenario includes two to three specific programming environments crafted by He et al., eliminating low-quality prompts and replicating diverse daily code completion tasks, making it a robust tool for assessing models' code security capabilities. These scenarios, based on incomplete code prompts in C/C++ or Python, challenge the models to produce appropriate code completions, highlighting their ability to handle incomplete inputs in real programming environments.


\subsection{Performance of RefleXGen}

To ensure a fair comparison, we initially set the RAG content to empty, allowing RefleXGen to progressively generate content during testing. We tracked several metrics: Sec. Rate , Pass Rate, Eff. Total,Sec. Count,Unres. Count. To obtain reliable data, we conducted five repeated experiments for each model. CodeQL \cite{szabo2023incrementalizing} was utilized to perform security analysis and assessment. In each experiment, every scenario was subjected to 25 task generations to average the results, ensuring an objective assessment of each model's generative capabilities.

\begin{figure}[!ht]
    \centering
    \includegraphics[width=0.8\linewidth]{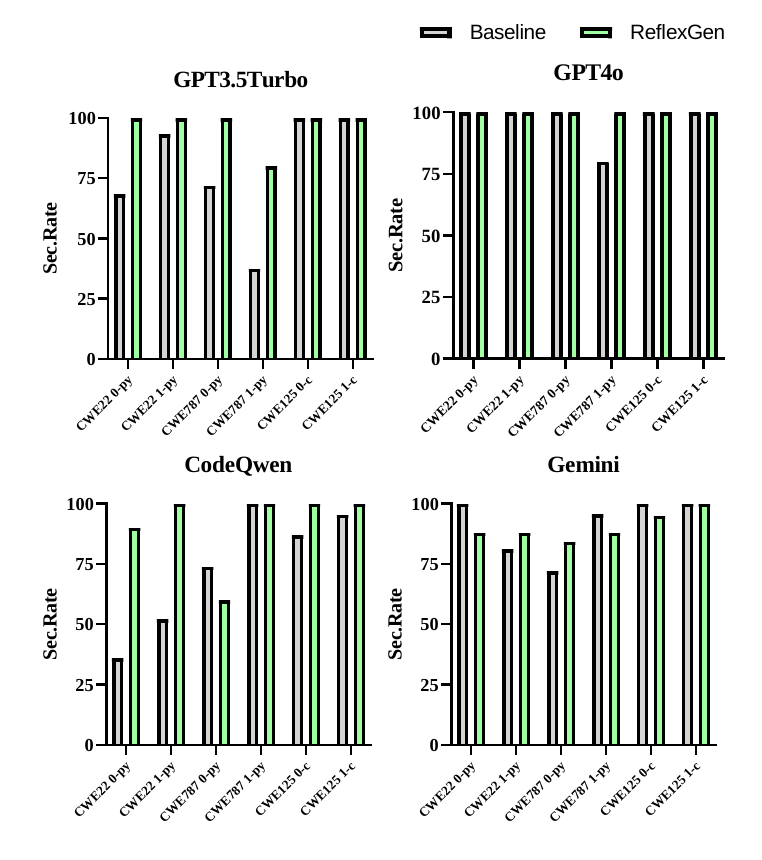}
    \caption{Sec.Rate Difference among Cases of RefleXGen}
    \label{fig:CWE-PIC}
\end{figure}

As shown in Table~\ref{table1}, RefleXGen demonstrated outstanding performance across four major models, effectively enhancing code security. Specifically, OpenAI's GPT-3.5 Turbo showed a 13.6\% improvement in code safety, GPT-4 improved by 6.7\%, CodeQwen-1.5 by 4.5\%, and Gemini-1.0-pro achieved a 5.8\% increase in security. These results indicate that the RefleXGen method significantly reduces the rate of defects and problematic code generation across different models.


Furthermore, we conducted a detailed analysis of three CWE scenarios that are typically concealed yet pose severe risks. As illustrated in Figure \ref{fig:CWE-PIC}, under the RefleXGen method, the code security generated by GPT-3.5 and CodeQwen demonstrated significant improvements in scenarios prone to triggering high-risk vulnerabilities. In contrast, Gemini exhibited fluctuations in security enhancements, while the improvements in GPT-4 were relatively modest, likely due to its already high baseline of code safety.


It is worth noting that, except for GPT-4, the initial compilation success rate for other models declined. This decline is primarily attributed to the introduction of more restrictive conditions and code interferences, which added complexity to the tasks. These changes led to more complex code outputs, thereby affecting the compilation success rates. However, GPT-4, with its robust overall capabilities, was less affected and even showed an improvement in compilation success. In contrast, CodeQwen, which has a smaller parameter size, experienced a greater decline. This phenomenon underscores the dependency of RefleXGen's enhancements on the models' capabilities in dialogue and handling complex scenarios.

\label{sec:typestyle}


%

\section*{CONCLUSION}



In this work, we have introduced RefleXGen, an innovative method that significantly enhances the security of code generated by large language models without the need for model fine-tuning or the creation of specialized security datasets. Universally applicable to all code generation models and operating independently of external enhancements, RefleXGen leverages the models' inherent reflective processes to accumulate security knowledge. By building a dynamic knowledge base, it optimizes prompts for subsequent code generation cycles. Experimental results demonstrate that RefleXGen substantially improves code generation security across various models, including GPT-3.5, GPT-4, CodeQwen, and Gemini, with particularly notable enhancements in models possessing stronger overall capabilities. This advancement underscores the potential of self-reflective mechanisms in AI models to autonomously improve code security, paving the way for future research in secure code generation without extensive resource investment.

\bibliographystyle{IEEEbib}
\bibliography{icassp} 

\end{document}